\documentclass{pazh_engl}
\usepackage{flushrt}
\usepackage{graphicx}
\usepackage{lscape}

\begin{document}

\title{\bf INTEGRAL/IBIS survey of the Sagittarius Arm Tangent region: A source catalog}

\author{\bf \hspace{-1.3cm}\copyright\, 2004 \ \
S.V.Molkov\inst{1}, A.M.Cherepashchuk\inst{2}, A.A.Lutovinov\inst{1}, 
M.G.Revnivtsev\inst{1,3}, K.A.Postnov\inst{2}, R.A.Sunyaev\inst{1,3}}

 \institute{Space Research Institute, Moscow, Russia
\and
Sternberg Astronomical Institute, Moscow, Russia
\and
Max-Plank-Institut fuer Astrophysik, Garching, Germany
}
\authorrunning{MOLKOV ET AL.}
\titlerunning{INTEGRAL/IBIS SURVEY OF SGR ARM TANGENT}

\date{Feb.16 2004}

\abstract{
Analysis of 18-120 keV images of the Sagittarius Arm Tangent region (SATR) obtained by  IBIS telescope onboard INTEGTRAL observatory during the spring of 2003 is performed. In the 18-60 keV energy range, 28 sources have been detected 
with a flux level above 1.4 mCrab. Of these sources, 16 were identified earlier as galactic X-ray binary systems, 3 as extragalactic objects, 2 as pulsars inside supernova remnants, and 7 has unknown nature. The analysis revealed the presence of three previously unknown sources. Fourteen sources show significant flux in the 60-120 keV energy range.}
\maketitle

\section*{Introduction}

Sagittarius Arm is the next toward the Galactic center spiral arm
of the Milky Way, $\sim 2$ kpc away from the Sun.  
Sagittarius arm is very rich in young massive stars and 
remnants of their evolution (high-mass X-ray binaries, microquasars,
X-ray pulsars, supernova remnants etc.). 

A set of very interesting X-ray 
sources, including the brightest microquasar 
GRS1915+105, the peculiar object SS433 and other black hole candidates (X1908+094, 
X1901+014), soft gamma-ray repeaters, supernova 
remnants, about dozen of 
persistent and transient X-ray bursters (e.g. 4U1915-05, Ser X-1, Aql X-1)
and X-ray pulsars (1855-026, 1907+097, GS 1843+009, X1901+03)
fall within SATR. 
Some of the sources in this field 
were discovered by INTEGRAL during the AO-1 observations.

Observations of this field and the Galactic Center regions during the
INTEGRAL AO-1 demonstrated excellent capabilites of the INTEGRAL telescopes
to construct sensitive maps, to perform good 
surveys and monitor transient sources 
in the field of view (Cherepashchuk et al. 2003, Molkov et al. 2003).

This paper is a part of work on hard X-ray cartography of our Galaxy and search for new point-like sources based on the IBIS data of the INTEGRAL observatory. First results on the deep Galactic Center survey are published elsewhere (Revnivtsev et al. 2004).

\section*{Observations and data analysis} 

The international gamma-ray observatory INTEGRAL was launched by the Russian launcher PROTON from the Baikonur cosmodrome in the high-apogee orbit on October 17, 2002 (Eismont et al. 2003). The payload includes four principal instruments which allow simultaneous observations of sources in the X-ray, gamma-ray and optical energy range (Winkler et al. 2003). The cartography and detection of point-like hard X-ray sources can be best performed with the top-layer detector ISGRI of the IBIS telescope (Lebrun et al. 2003). This telescope uses the coded aperture method and allows imaging of a given sky area within a field of view of   $29^\circ\times29^\circ$ (the fully coded area is $9^\circ\times9^\circ$) in the 15-200 keV range with an angular resolution of $12^\prime$.

The SAT field have regularly been observed by INTEGRAL during the first year of operation in the orbit. In the present work we used data obtained in the TOO observations of the Aql X-1 flare in March-April 2003 (Molkov et al. 2003)
and SS433 in May 2003 (Cherepashchuk et al. 2003a). The total exposure in these observations amounts to  
$\sim 830$ ks. The dithering mode used during the observations (24 pointings 2 and 4 degrees off the central source) enables recovering the image of a substantial sky area $35^\circ\times40^\circ$ in this part of the Galaxy.

The data of all observations were processed using methods described in Revnivtsev et al. (2004) paper. Analysis of an extensive set of calibration observations of the Crab nebula with different location of this sources within the IBIS FOV suggests that with the approach and software employed the conservative estimation of uncertainty in measurements of the absolute fluxes from the sources is about 10\%. The localization accuracy of bright Crab-like sources is  
$\sim$0.4$^\prime$ ($1\sigma$) decreases to 
$\sim$2-3$^\prime$ for weak sources (with the signal-to-noise ratio
$\sim 5-6$, 
see below). 

The presence of a peak above some threshold was considered to be a signature of a point-like source. The statistical threshold was chosen after the entire image had been analyzed. The signal-to-noise ratio ($S/N$) that ensures registration of at most one false source in the full studied field was derived. 

Fig. 1 shows the signal-to-noise ratio distribution for all sources found in the full image of the region in a logarithmic scale. For an ideal image with negligibly small systematical deviations this distribution in the region without sources should be described by the normal law with zero mathematical expectation and unit variance.  
In Fig. 1 such a distribution is shown by the solid curve in the logarithmic scale. Fig. 1 clearly indicates that the negative part of the actually obtained $S/N$ distribution is well fitted by the theoretically expected one down to values of order of $S/N\sim -4$. Note also that small deviations in the range $-5 < S/N<-4$ are also seen, with the measured distribution showing a slightly larger number of points than theoretical one. So we can conclude that it would be incorrect to utilize purely theoretically calculated value of the source registration threshold ($\sim 4.2 \sigma$). Therefore, the lower threshold of the confident source detection was chosen to be ($>5 \sigma$). 
\begin{figure}[htb]
\includegraphics[width=\columnwidth]{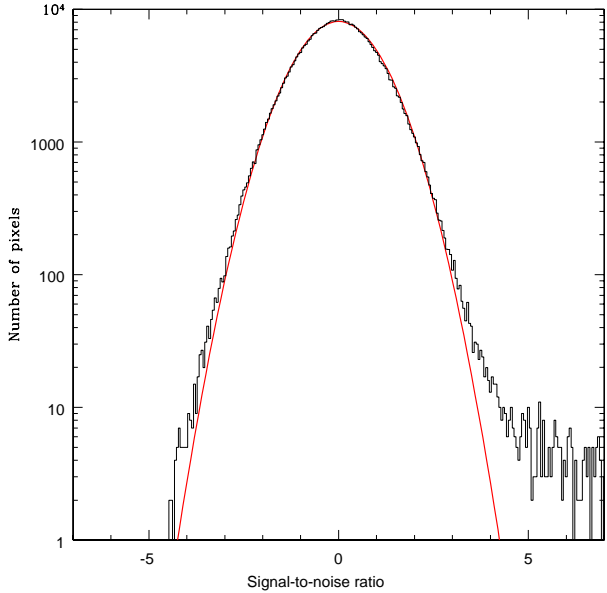}
\caption{The signal-to-noise ratio distribution for all points of the image of the studied region. The solid curve shows the normal distribution with zero mathematical  expectation and unit variance}
\end{figure}

\section*{Results}

Table 1 shows the list of all registered sources with their coordinates and fluxes in units of mCrab in two energy ranges. A flux of 1 mCrab in the 18-60 keV and 60-120 keV energy ranges from a source with power-law spectrum and photon index $\Gamma=2.1$ corresponds to energy fluxes  
$\sim1.4\times 10^{-11}$ erg/s/cm$^2$ and $\sim7.1\times 10^{-12}$
erg/s/cm$^2$, respectively. The localization accuracy of the sources is $\sim$2--3$^\prime$ (the 90\% radius of the confidence contour). The sources are listed in the order of decreasing detection confidence, which is dependent upon both their intensity and the effective observation time of each object.    

The 18-60 keV image of the Sagittarius Arm Tangent region with a size of $\sim 35^\circ \times 25^\circ$ is shown in Fig. 2. in galactic coordinates. The analysis performed allowed us to detect 28 sources in the 18-60 keV energy band, part of which (14) persists sufficiently bright to be detected in harder 60-120 keV energy range (Table 1). We repeat that sources are listed in the order of their detection confidence, which is determined not only by the intrinsic source flux level but also by the effective exposure time.

\begin{figure*}[t]
\includegraphics[width=0.65\textheight]{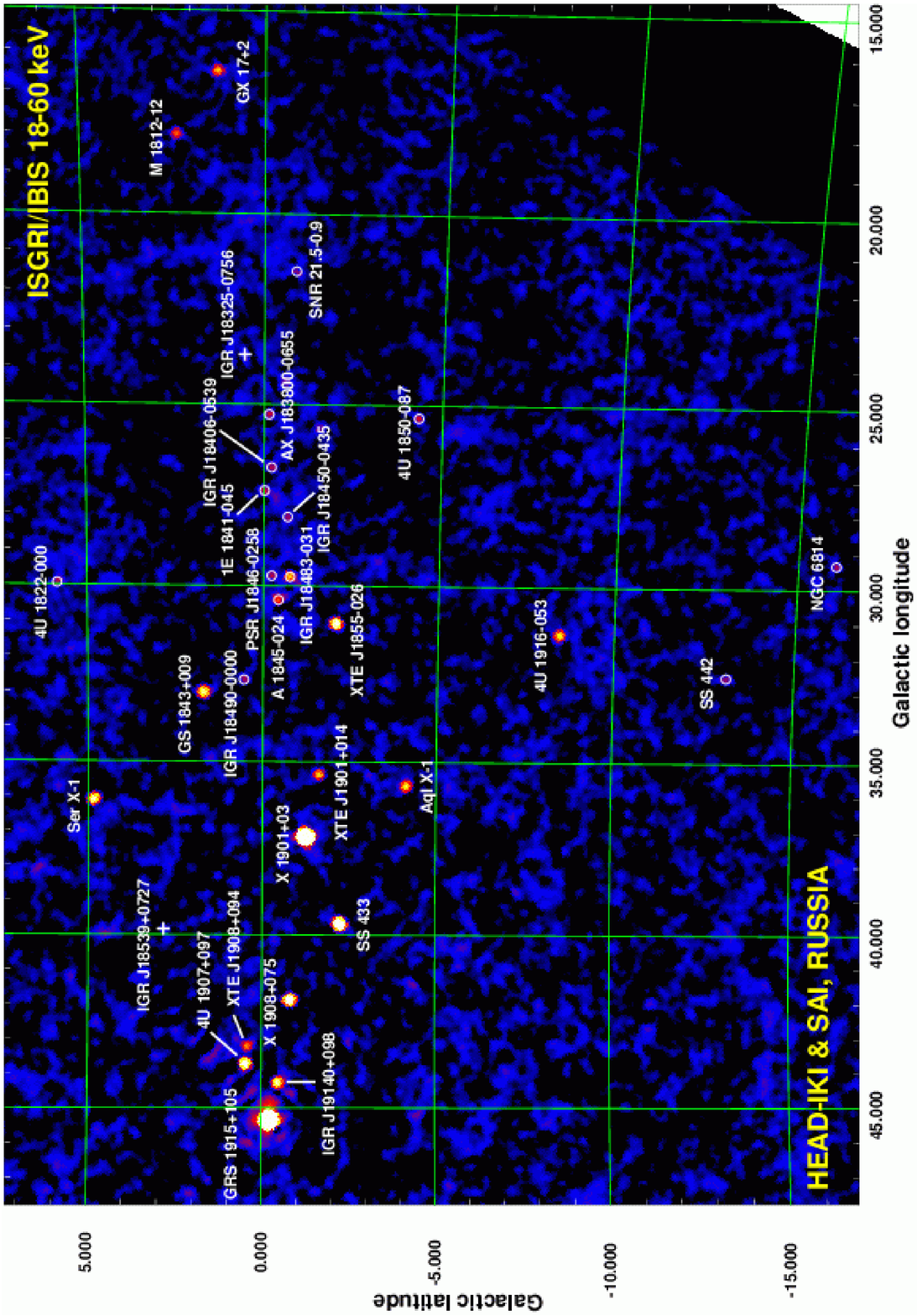}
\caption{The mosaic image of the Sagittarius Arm Tangent region obtained by the INTEGRAL IBIS telescope in the energy range 18-60 keV. The total exposure time is  $\sim 800$ ks.}
\end{figure*}

Some of the detected sources (21 of 28) are of the known nature and either belong to high-mass X-ray binaries (HMXB) or low mass X-ray  binaries (LMXB). In Table 1, the classification of the registered sources is shown in the last column using the following notations: LMXB -- low mass X-ray binaries with the optical companion mass  $\sim 1M_\odot$; HMXB --  high mass X-ray binaries with the optical companion mass $\geq 8M_\odot$; BH --  black hole candidates; T -- transient sources; P -- X-ray pulsars, B -- X-ray bursters, Z -- $Z$-sources; G -- sources in globular clusters; AXP -- anomalous X-ray pulsars, SNR -- supernova remnants; AGN -- active galactic nuclei. 

Four black hole candidates were confidently detected during observations, including the well-known microquasar GRS 1915+105 (Hannikainen et al. 2003), 
SS433 (Cherepashchuk et al. 2003), 
XTE~J1908+094 and X~1908+075 (Wen et al. 2000).

In addition, a significant X-ray flux was detected from seven 
X-ray pulsars, including X~1901+03,
4U~1907+097, XTE~J1855-026, GS~1843+009, A~1845-024,
1E~1841-045 (anomalous X-ray pulsar) and PSR~J1846-0258.
The first of them (X~1901+03) has been known earlier only as a transient high mass binary (Forman et al. 1976), which was in inactive state since 1971. In the beginning of 2003, the source became active again and coherent X-ray pulsations were found (Galloway et al. 2003a,b; Molkov et al. 2003). 

Of seven detected low mass X-ray binaries with neutron stars six are X-ray bursters, with one of them residing in globular cluster NGC~6712.  

During the observations, significant X-ray fluxes were also detected from three supernova remnants and two active galactic nuclei (NGC~6814 and SS~442/1H1934-063). The identification of the last source is ambiguous. The point is that the location of the source as derived from our analysis is different from more accurate location SS~442/1H1934-063 taken from catalog by about $\sim
5.6^\prime$. However, as the source is only marginally detected ($\sim 5 \sigma$) and the 90\% level of its localization contour is 
$\sim 3 ^\prime$, there is the $\sim$10\% probability that its location differs from the real one by more 3$^\prime$. Nonetheless, it can not be excluded that this is the new source IGR~J19378-0617. 

Two known X-ray sources discovered by RXTE 
 (XTE~J1901+014) and ASCA (AX~J183800-0655) were reliably registered in our observations. As yet, their nature remains unclear. 

Finally, seven new INTEGRAL sources fell within the IBIS telescope FOV. Of these sources, five are confidently detected in our observations: IGR~J19140+098 (Hannikainen et al. 2003b), IGR~J18483-031 (Chernyakova et al. 2003), IGR~J18490-0000,
IGR~J18406-0539, IGR~J18450-0435 (the last three were first discovered in the present work). The remaining two sources 
IGR~J18325-0756 and IGR~J18539+0727, discovered earlier by the INTEGRAL observatory (Lutovinov et al. 2003a,b), were not registered in our observations and we provide here only upper limits of their fluxes. Note that the IGR~J18539+0727 is apparently a black hole candidate (Lutovinov and Revnivtsev 2003). 

To conclude, of 28 sources detected during the INTEGRAL AO-1 Saggitarius Arm Tangent region observations, seven objects are of unknown nature, with five of which are first discovered by the INTEGRAL observatory.

\bigskip

Authors thank Eugene Churazov for developing of algorithms of 
analysis of data of IBIS telescope and for providing the software.
The work is supported by the Ministerly of Industry and Science of Russia (the President of RF grant No NSh-2083.2003.2) and the Rusian Academy of Sciences
''Non-stationary phenomena in astrophysics''. KAP acknowledges the RFBR grant No 03-02-16110. The authors acknowledge The INTEGRAL Science Data Center (Versoix, Switzerland) and INTEGRAL Russian Scientific Data Center (Moscow, Russia). 
The work is based on observations with INTEGRAL, an ESA project with
instruments and science data centre funded by ESA member states
(especially the PI countries: Denmark, France, Germany, Italy,
Switzerland, Spain), Czech Republic and Poland, and with the
participation of Russia and the USA. 

\section*{REFERENCES}

\parindent=0mm 

Cherepashchuk A.M., Sunyaev R.A., Seifina E. et al.
//  Astron. Astroph. {\bf 411}, L441, (2003Á)

Cherepashchuk, A. M., Molkov, S., Foschini, L. et al.
//  The Astronomer's Telegram, 159 (2003Â)

Chernyakova M., Lutovinov A., Capitanio F., et. al.//
Astron. Telegramm ATEL {\bf 157} (2003)

Eismont N.A., Ditrikh A.V.,  Janin G. et al.//
Astron. Astroph. {\bf 411}, L37 (2003) 

Forman W., Tananbaum H., Jones C.// Astroph. J.
{\bf 205}, L29 (1976)

Galloway D., Remillard R., Morgan E.//
IAUC 8070, (2003a)

Galloway D., Remillard R., Morgan E.//
IAUC 8081, (2003b)

Hannikainen D. C., Vilhu O., Rodriguez J. et al.
//  Astron. Astroph. {\bf 411}, L415, (2003a)

Hannikainen, D. C.; Rodriguez, J.; Pottschmidt,
K. et al. // IAU Circ., 8088, (2003b)

Lebrun F., Leray J. P., Lavocat P. et al.//
Astron. Astroph. {\bf 411}, L141 (2003) 

Lutovinov A., Shaw S., Foschini L., et. al.) Astron.
Telegramm ATEL {\bf 154} (2003a) 

Lutovinov A., Rodriguez J., Produit N., et. al.)
Astron. Telegramm ATEL {\bf 151} (2003b)

Lutovinov A., Revnivtsev M.// Astronomy Lett. {\bf 29}, 719 (2003)

Molkov S.V., Lutovinov A.A., Grebenev S.A. //
Astron. Astroph. {\bf 411}, L357 (2003) 

Revnivtsev M., Sunyaev R., Varshalovich D.// Astronomy Lett. in press (2004); 
astro-ph/0402027 

Wen L., Remillard R. A., Bradt H. V. // Astroph. J. {\bf
  532}, 1119 (2000) 

Winkler C., Courvoisier T. J.-L., Di Cocco G. et al.//
Astron. Astroph. {\bf 411}, L1 (2003)

\begin{table*}
\caption{List of sources, detected during observations of Sgr Arm tangent
region in March-May 2003}
\begin{tabular}{l|c|c|c|c|l|l}
\hline
&\multicolumn{2}{|c|}{J2000}& \multicolumn{2}{|c|}{{~~~~Mean flux, mCrab $^a$~~~~}}  & ~~~~~Identification~~~ & ~~~~Class~~\\
\cline{2-5}
&~~~~~~~$\alpha$~~~~~~~&~~~~~~$\delta$~~~~~~ & ~~~~~18-60 keV~~~~~ & ~~~~~60-120 keV~~~~~  & \\
\hline
  1 &   288.80 &    10.95 & $  252.6 \pm     0.2$ & $   80.0 \pm  0.7$ & {GRS~1915+105}                & LMXB,~BH \\
  2 &   285.91 &     3.21 & $   77.4 \pm     0.2$ & $    3.2 \pm  0.6$ & {X~1901+03 $^b$}              & HMXB,~TP \\
  3 &   287.96 &     4.99 & $   14.0 \pm     0.2$ & $    6.2 \pm     0.6$ & SS~433    $^c$              & HMXB,~BH \\
  4 &   287.71 &     7.60 & $   10.9 \pm     0.2$ & $    6.8 \pm     0.6$ & X~1908+075            & HMXB,~BH? \\
  5 &   287.42 &     9.83 & $   10.9 \pm     0.2$ & $  0.8 \pm 0.7$ $^h$ & 4U~1907+097                 & HMXB,~TP \\
  6 &   283.87 &    -2.60 & $    8.8 \pm     0.2$ & $    4.4 \pm     0.7$ & XTE~J1855-026               & HMXB,~P \\
  7 &   279.99 &     5.03 & $    7.5 \pm     0.2$ & $  0.6 \pm  0.7$ $^h$ & Ser~X-1                     & LMXB,~B \\
  8 &   288.53 &     9.87 & $    6.5 \pm     0.2$ & $    2.8 \pm     0.7$ & IGR~J19140+098              & ? \\
  9 &   281.40 &     0.86 & $    6.3 \pm     0.2$ & $    3.6 \pm     0.7$ & GS~1843+009 $^b$            & HMXB,~TP \\
 10 &   287.79 &     0.56 & $    4.1 \pm     0.2$ & $  1.3 \pm 0.6$ $^h$ & Aql~X-1 $^b$                 & LMXB,~TBA \\
 11 &   289.69 &    -5.24 & $    5.2 \pm     0.2$ & $  1.4 \pm 0.8$ $^h$ & 4U~1916-053                 & LMXB,~BD \\
 12 &   285.41 &     1.45 & $    3.6 \pm     0.2$ & $    3.1 \pm     0.6$ & XTE~J1901+014               & ? \\
 13 &   282.10 &    -3.16 & $    4.3 \pm     0.2$ & $    3.9 \pm     0.8$ & IGR~J18483-031              & ? \\
 14 &   274.03 &   -14.02 & $   42.9 \pm     2.4$ & $  3.2 \pm 8.1$ $^h$ & GX~17+2                     & LMXB,~ZB \\
 15 &   287.22 &     9.35 & $    2.9 \pm     0.2$ & $    4.4 \pm     0.6$ & XTE~J1908+094               & HMXB,~BH \\
 16 &   273.77 &   -12.12 & $   20.6 \pm     1.7$ & $   24.0 \pm     5.8$ & M~1812-12                   & LMXB,~B \\
 17 &   282.08 &    -2.48 & $    2.7 \pm     0.2$ & $  1.7 \pm 0.9$ $^h$ & A~1845-024                  & HMXB,~TP \\
 18 &   283.25 &    -8.70 & $    2.6 \pm     0.3$ & $  3.7 \pm     1.1  $ & 4U~1850-087        & LMXB,~GB \\
 19 &   279.50 &    -6.91 & $    2.7 \pm     0.4$ & $  1.8 \pm 1.1$ $^h$ & AX~J183800-0655 & ? \\
 20 &   280.37 &    -4.94 & $    2.1 \pm     0.3$ & $  6.8 \pm  0.9$ & 1E~1841-045 & AXP,~SNR \\
 21 &   281.61 &    -2.97 & $    1.6 \pm     0.2$ & $  3.4 \pm  0.8$ & PSR~J1846-0258 & P,~SNR \\
 22 &   295.60 &   -10.36 & $    4.1 \pm     0.6$ & $  4.6 \pm 2.1$ $^h$ & NGC~6814 & AGN \\
 23 &   282.25 &    -0.00 & $    1.4 \pm     0.2$ & $ 0.6 \pm 0.7$ $^h$ & IGR~J18490-0000 $^c$         & ? \\
 24 &   276.32 &    -0.03 & $    2.2 \pm     0.3$ & $ 0.5 \pm 1.1$ $^h$& 4U~1822-000                 & LMXB \\
 25 &   280.23 &    -5.65 & $    2.0 \pm     0.3$ & $ 1.7 \pm 1.0$ $^h$ & {IGR~J18406-0539 $^c$}     & ? \\
 26 &   281.25 &    -4.58 & $    1.5 \pm     0.3$ & $ 1.7 \pm 0.9$ $^h$ & IGR~J18450-0435 $^c$          & ? \\
 27 &   294.46 &    -6.28 & $    1.7 \pm     0.3$ & $ 1.1 \pm 1.1$ $^h$ & SS~442/1H1934-063 $^d$ & AGN\\
 28 &   278.40 &   -10.58 & $    3.1 \pm     0.6$ & $ 4.6 \pm 2.1$ $^h$ & SNR~21.5-0.9    & SNR \\
 29 &   278.12 &    -7.93 & $    1.6 \pm     0.5$ & $ 1.1 \pm 1.5$ $^h$& IGR~J18325-0756 $^e$        & ? \\
 30 &   283.48 &     7.45 & $    0.6 \pm     0.2$ & $ 0.5 \pm 0.7$ $^h$ & IGR~J18539+0727 $^e$        & BH? \\
\hline
\end{tabular}

$^a$ --- only statistical errors are given (systematical error is $\sim 10\%$); \\
$^b$ --- sources observed during flares  (Molkov et al 2003; Cherepashchuk et al. 2003 a,b)); \\
$^c$ --- sources discovered in the present observations;\\
$^d$ --- the observed location of this source differs from more accurate localization of SS~442 by $\sim5.6'$; possibly this is a new object
IGR~J19378-0617 (see the text); \\
$^e$ --- sources discovered by the IBIS telescope in other INTEGRAL observations; \\
$^h$ --- upper limits for fluxes of these sources 
\end{table*}

\end{document}